\documentclass[twocolumn,secnumarabic,amssymb, nobibnotes, aps, prd, showpacs]{revtex4-1}
\usepackage{amsmath}
\usepackage{graphicx}
\usepackage{amssymb}
\usepackage{subfigure}
\usepackage{color}
\usepackage{lmodern}
\usepackage{lineno}

\begin{document}
% Use the \preprint command to place your local institutional report
% number in the upper righthand corner of the title page in preprint mode.
% Multiple \preprint commands are allowed.
% Use the 'preprintnumbers' class option to override journal defaults
% to display numbers if necessary
%\preprint{}

%\linenumbers

%Title of paper
\title{Brownian Thermal Noise in Functional Optical Surfaces}

% repeat the \author .. \affiliation  etc. as needed
% \email, \thanks, \homepage, \altaffiliation all apply to the current
% author. Explanatory text should go in the []'s, actual e-mail
% address or url should go in the {}'s for \email and \homepage.
% Please use the appropriate macro foreach each type of information

% \affiliation command applies to all authors since the last
% \affiliation command. The \affiliation command should follow the
% other information
% \affiliation can be followed by \email, \homepage, \thanks as well.
\author{S.~Kroker}
\affiliation{Physikalisch-Technische Bundesanstalt, Bundesallee 100, 38116 Braunschweig, Germany}
\affiliation{Technische Universit\"at Braunschweig, LENA Laboratory for Emerging Nanometrology, Pockelsstra{\ss}e 14, 38106 Braunschweig, Germany}
\email[]{stefanie.kroker@ptb.de}
\author{J.~Dickmann, C.~B.~Rojas~Hurtado}
\affiliation{Physikalisch-Technische Bundesanstalt, Bundesallee 100, 38116 Braunschweig, Germany}
\author{D.~Heinert, R.~Nawrodt}
\affiliation{Friedrich-Schiller-Universit\"at Jena, Institut f\"ur Festk\"orperphysik, Helmholtzweg 5, 07743 Jena, Germany}
\author{Y.~Levin}
\affiliation{School of Physics and Astronomy, Monash University, PO Box 27, VIC 3800, Australia}
\author{S.~P.~Vyatchanin}
\affiliation{Faculty of Physics, Moscow State University, Moscow 119991, Russia}

%\homepage[]{Your web page}
%\thanks{}
%\altaffiliation{}
%\affiliation{}

%Collaboration name if desired (requires use of superscriptaddress
%option in \documentclass). \noaffiliation is required (may also be
%used with the \author command).
%\collaboration can be followed by \email, \homepage, \thanks as well.
%\collaboration{}
%\noaffiliation

\date{\today}

\begin{abstract}
We present a formalism to compute Brownian thermal noise in functional optical surfaces such as grating reflectors, photonic crystal slabs or complex metamaterials. Such computations are based on a specific readout variable, typically a surface integral of a dielectric interface displacement weighed by a form factor. This paper shows how to relate this form factor to Maxwell's stress tensor computed on all interfaces of the moving surface. As an example, we examine Brownian thermal noise in monolithic T-shape grating reflectors. The previous computations by Heinert \textit{et al.} [Heinert \textit{et al.}, PRD 88 (2013)] utilizing a simplified readout form factor produced estimates of thermal noise that are tens of percent higher than those of the exact analysis in the present paper. The relation between the form factor and Maxwell's stress tensor implies a close correlation between the optical properties of functional optical surfaces and thermal noise.
\end{abstract}

% insert suggested PACS numbers in braces on next line
% insert suggested keywords - APS authors don't need to do this
%\keywords{}

%\maketitle must follow title, authors, abstract, \pacs, and \keywords
\maketitle

% body of paper here - Use proper section commands
% References should be done using the \cite, \ref, and \label commands
\section{Introduction}
\begin{figure}[t]
	\centering
		\includegraphics[width=0.40\textwidth]{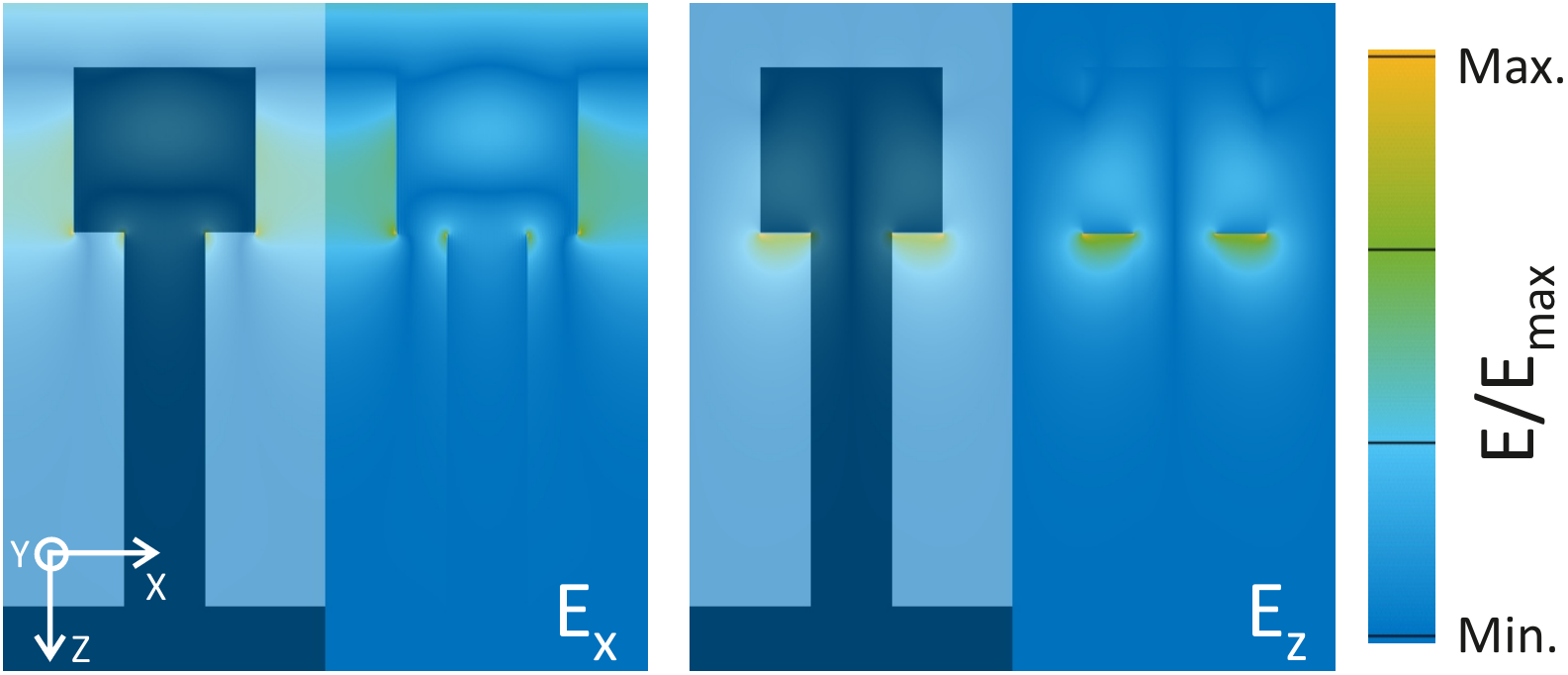}
			\caption{Distribution of $E_\mathrm{x}$ and $E_\mathrm{z}$ in the high reflectivity grating reflector discussed by Heinert \textit{et al.} \cite{Hei2013}. The calculation was performed by means of the rigorous coupled wave algorithm (RCWA) \cite{Moh1981} for a wavelength of 1550\,nm, normal incidence and transverse-magnetic polarized light.}
	\label{fig:Tshapefield}
\end{figure}

Thermal noise sets a crucial limitation to several high precision instruments, for example ultra-stable laser resonators for the realization of optical clocks, high resolution optical spectroscopy and gravitational wave detectors \cite{Sau1990,Num2004,Kes2012, Zha2014, Hag2014, Gra2017}. Particularly, Brownian displacement noise from random motion of amorphous optical coatings, as utilized for high reflectivity Bragg mirrors, represents a severe bottleneck for future sensitivity improvements of these measurement systems \cite{Lev1998, Har2002, Pen2003, Hil2011, Hon2013}. The reason for the large Brownian noise amplitude is the high mechanical loss of the coating materials. Currently, several approaches to reduce Brownian coating thermal noise are under investigation, for example optimizing the mechanical properties of amorphous materials, or using crystalline coating stacks based on AlGaAs/GaAs and AlGaP/GaP as low-loss coating materials \cite{Har2006,Pri2015, Col2013, Cum2015, Lin2015, Gra2016}. \newline
As an alternative to Bragg mirrors, grating reflectors based on crystalline silicon have been theoretically proposed \cite{Bru2008} and experimentally realized \cite{Bru2010, Kro2013, Kro2013a}. Since these elements can be monolithically implemented without adding any amorphous material with high mechanical loss, they are promising as low-noise optical components. In contrast to Bragg mirrors, in grating reflectors high reflectivity is realized by an optical resonance which leads to a penetration of the light into a surface layer of only a few hundred nanometers thickness \cite{Lal2006,Kar2012}. Fig.~\ref{fig:Tshapefield} illustrates a typical field distribution in a monolithic high reflectivity structure. The lower grating region acts as a supporting structure that prevents the light from leaking into the substrate.\newline
A typical task in high-precision opto-mechanical experiments is to measure the phase shift of light reflected from a mirror surface - or, alternatively, the change of the optical mode frequency if the mirror is a part of the optical resonator. For small displacements, this readout variable depends linearly on the displacement of the reflecting surface. It can be expressed as:
\begin{equation}
\hat{z}(t)=\int f(\vec{r})u_\mathrm{z}(\vec{r},t)\ \mathrm{d}A.
\label{Eq:readout}
\end{equation}

\noindent$\vec{r}$ is the location of a point on the surface, and $u_\perp(\vec{r},t)$ is the displacement of the mirror perpendicular to the surface $A$ at $\vec{r}$ and time $t$. The form factor $f(\vec{r})$ depends on the intensity profile of the laser beam and is proportional to the laser light intensity at $\vec{r}$ as will be shown below. 
In the case of a planar surface the form factor is simply the laser beam profile whereas in the case of a structured surface the determination of $f(\vec{r})$ is a non-trivial task. The standard way to compute thermal noise in this variable is to use a formulation of the fluctuation-dissipation theorem by Callen and Welton~\cite{Cal1951} which employs a virtual oscillating pressure of the form \cite{Lev1998}:

\begin{equation}
p(\vec{r})=F_0\cos(\omega t)f(\vec{r}),
\label{Eq:pressure}
\end{equation}

\noindent where $f(\vec{r})$ is the form factor of Eq.~\ref{Eq:readout}. The virtual pressure is utilized to determine the strain energy density $\epsilon(x,y,z)$. This strain distribution then serves as a basis to calculate the dissipated mechanical energy in the system at a given frequency $\omega$. Using the model of structural loss, the dissipated energy reads \cite{Lev1998}:
\begin{equation}
W_\mathrm{diss}(\omega)=\omega \int{\epsilon(\vec{r})\phi(\vec{r})\ \mathrm{d}V},
\label{Eq:Wdiss}
\end{equation}
\noindent where integral needs to be performed over the whole component under investigation.
The Brownian thermal noise power spectral density can be expressed by:

\begin{equation}
S_\mathrm{z}(\omega, T)=\frac{8k_\mathrm{B}T}{\omega^2}\frac{W_\mathrm{diss}}{F_0^2}.
\label{Eq:noise}
\end{equation}

\noindent The challenge is to compute the form factor on arbitrary surfaces, and this paper gives a direct and exact answer. The previous approach by Heinert \textit{et al.} \cite{Hei2013} gave an approximation by assuming, that the form factor was constant on large segments on the interface. Heinert \textit{et al.} evaluated the impact of displacing these segments as a whole on the overall phase shift of the reflected light. In contrast, this paper finds that the form factor is strongly inhomogeneous, which significantly affects the computation of thermal noise spectral density. \newline
As an application of our formalism, we investigate Brownian thermal noise in monolithic silicon T-shape grating reflectors and compare the results with the work by Heinert \textit{et al.} \cite{Hei2013}. In addition, we investigate the impact of width of the support structure as a critical parameter for Brownian thermal noise. We find that an optimum support structure width exists which minimizes Brownian noise. Due to manufacturing errors, the geometric dimensions of the grating may differ from the design values by a few nanometers. We evaluate the consequences of manufacturing errors and show that it may lead to deviations of thermal noise by a factor of about 2.5.
\newline
The article is organized as follows: In Sec.~\ref{sec:calcul} we introduce the calculation method based on Maxwell's stress tensor. Afterwards, in Sec.~\ref{sec:params} we discuss how the geometric grating parameters of T-shape grating reflectors with different support structure widths were defined. In Sec.~\ref{sec:results} we utilize these parameters to compute the virtual forces required for the thermal noise calculations, the energy of elastic deformation in response to these forces, and finally the Brownian thermal noise. 

\section{Calculation of Brownian thermal noise in functional optical surfaces}
\label{sec:calcul}

\noindent The form factor can be described following \cite{Dem2015, Tug2017}. Let us first consider an optical cavity of length $L$. When one of the two mirrors is moved by a displacement of $u$ the eigenfrequency $\omega$ of the cavity is changed by: 
\begin{equation}
\frac{\delta \omega}{\omega}=\frac{z}{L}.
\end{equation}

\noindent The quantity $z$ contains the measurement signal, e.g. a gravitational wave signal. But also random perturbations $u_z(\vec{r})$ caused by Brownian motion may contribute to a frequency change and thus disturb the measurement signal. The question to be answered is, how such a displacement translates into the frequency change of the cavity. A slow displacement does not change the number of photons in the cavity. This condition of adiabaticity is satisfied very well if the frequencies of interest are much smaller than the inverse light roundtrip inside the cavity, as is valid for the LIGO gravitational wave detector. In this case the relation

\begin{equation}
\frac{\mathcal{E}}{\omega}=\mathrm{const.}
\end{equation}
is fulfilled. Therefore, a change of the energy $\delta\mathcal{E}$ may be converted into frequency change of the optical eigenmode $\delta\omega$:

\begin{equation}
\frac{\delta \mathcal{E}}{\mathcal{E}}=\frac{\delta \omega}{\omega}.
\end{equation}

\noindent The energy change is a result of the work performed against the ponderomotive pressure perpendicular to the surface $p_{\perp}$. Thus, the energy change of the optical cavity mode caused by displacements $z_\mathrm{z}(\vec{r})$ can be expressed by:

\begin{equation}
\delta\mathcal{E}=\int p_\perp(\vec{r}) u_\mathrm{z}(\vec{r})\ \mathrm{d}A.
\end{equation}

\noindent The ponderomotive pressure relates a perturbation $u_\mathrm{z}(\vec{r})$ of an arbitrary surface to an effective translation $\hat{u}$ of the cavity mirror as a whole:

\begin{equation}
\hat{z} = \frac{L}{\mathcal{E}}\int p_\perp(\vec{r}) u_\mathrm{z}(\vec{r})\ \mathrm{d}A
\label{Eq:zhat}
\end{equation}

\noindent The ponderomotive light pressure results from the difference of Maxwell's stress tensor on both sides of the interface:

\begin{equation}
 p_{\perp}(\vec{r})=\Delta \sigma_{ij}(\vec{r})n_{i}n_{j},
\end{equation}
where $n_{i}$ is the unit vector normal to the surface and a summation over the dummy indices $i$ and $j$ is implied.    
\noindent Maxwell's stress tensor SI-units reads:

\begin{align}
	\sigma_{ij}= \varepsilon_0\varepsilon_\mathrm{r} E_i E_j + \frac{1}{\mu_0\mu_\mathrm{r}} B_i B_j - \frac{1}{2}\left(\varepsilon_0\varepsilon_\mathrm{r} E^2 + \frac{1}{\mu_0\mu_\mathrm{r}}B^2\right)\delta_{ij}.
	\label{Eq:Ediss}
\end{align}

\noindent $\varepsilon_0$ and $\mu_0$ are the dielectric and magnetic field constants, $E_i$ is the vacuum electric field amplitude and $B_i$ the magnetic field amplitude, respectively. On arbitrary surfaces, the electromagnetic field distribution can be calculated with the finite element tool COMSOL \cite{comsol}.
In the following sections, we will use the Maxwell stress tensor to evaluate the virtual forces in T-shaped monolithic grating reflectors and derive Brownian thermal noise thereof. Since the electric and magnetic fields depend on the position at the surface, the stress tensor components $\sigma_{ij}$ are also a function of the position. For the sake of readability, we will omit this explicit spatial dependency in our notation.

\section{Virtual pressure in monolithic T-shape grating reflectors}

\begin{figure}[tb]
	\centering
		\includegraphics[width=0.25\textwidth]{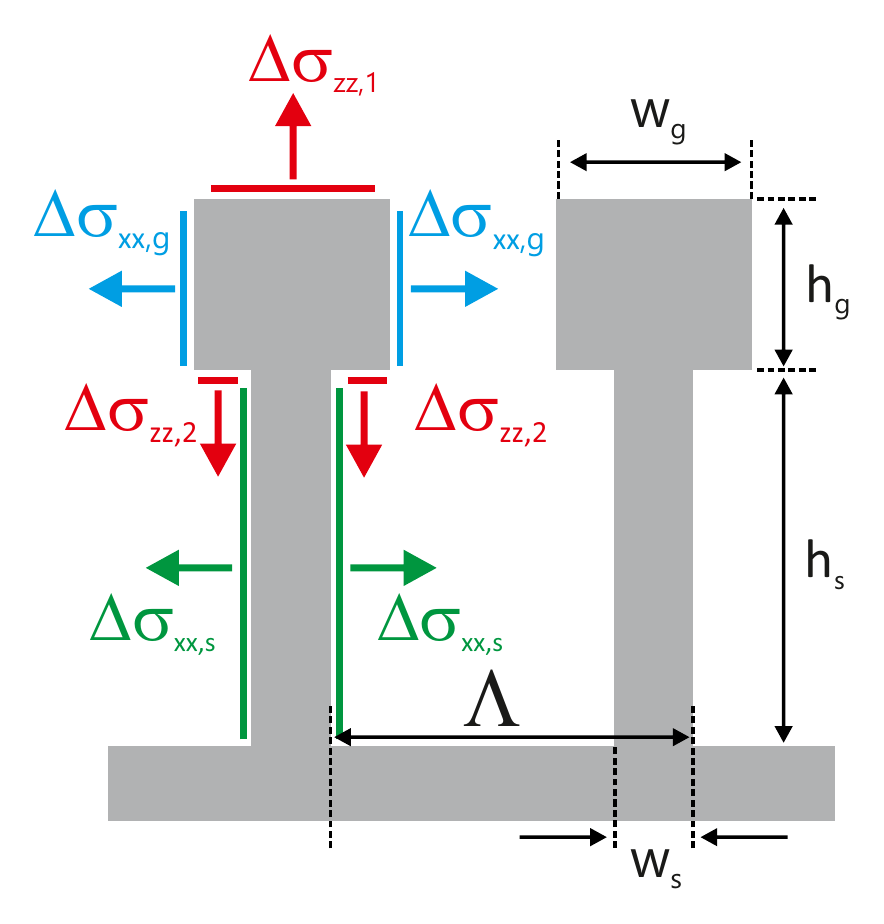}
	\caption{Distribution of the virtual pressure components (left). Overview of the geometric structure parameters (right).}
	\label{fig:sketch}
\end{figure}

\noindent In a T-shape structure the relevant components of the stress tensor are $\sigma_\mathrm{xx}$ and $\sigma_\mathrm{zz}$ (see Fig.~\ref{fig:sketch}) and the resulting pressure is the difference of the pressures inside and outside the structure. For non-magnetic materials ($\mu_\mathrm{r}=1$) the relevant pressure components at the grating surface are: 

\begin{align}
\Delta	\sigma_\mathrm{xx} &= \frac{\varepsilon_0}{2} (\varepsilon_\mathrm{r}-1) \left( \frac{E_\mathrm{x}^2}{\varepsilon_\mathrm{r}}+
E_\mathrm{y}^2+E_\mathrm{z}^2\right), \\
\Delta	\sigma_\mathrm{zz} &= \frac{\varepsilon_0}{2} (\varepsilon_\mathrm{r}-1) \left( \frac{E_\mathrm{z}^2}{\varepsilon_\mathrm{r}}+
E_\mathrm{y}^2+E_\mathrm{x}^2\right).
\label{Eq:maxwell}
\end{align}

\noindent $E_\mathrm{x}$, $E_\mathrm{y}$ and $E_\mathrm{z}$ are the vacuum fields and $\varepsilon_\mathrm{r}$ is the relative permittivity of the grating material. To relate our method to the results of Heinert \textit{et al.} \cite{Hei2013}, we will restrict our considerations on light with transverse magnetic (TM) polarization ($E_y=0$). In this case the pressure components reduce to:
	
\begin{align}
\Delta	\sigma_\mathrm{xx} &= \frac{\varepsilon_0}{2} (\varepsilon_\mathrm{r}-1) \left( \frac{E_\mathrm{x}^2}{\varepsilon_\mathrm{r}}+E_\mathrm{z}^2\right), \\
\Delta	\sigma_\mathrm{zz} &= \frac{\varepsilon_0}{2} (\varepsilon_\mathrm{r}-1) \left( \frac{E_\mathrm{z}^2}{\varepsilon_\mathrm{r}}+E_\mathrm{x}^2\right). 
	\end{align}
	
\noindent By using finite element analysis, the pressure components can be calculated and applied to the surface of the structure \cite{comsol}. Using Eq.~\ref{Eq:zhat}, one can show that $F_0$ is the overall radiation pressure force from the light beam onto the mirror. On can evaluate it directly from the Maxwell stresses at the dielectric interface. In this case one should carefully keep track of the sign contributions from the force applied at different segments of the grating as shown in Fig.~\ref{fig:sketch}. The resulting force $F_0$ is the integral of the pressure over the surface $A$ of a single period normalized to a unity length in $y$-direction parallel to the ridges. The elastic energy then is the volume integral of the energy density $\epsilon$ over one T-shape ridge. In combination with the mechanical loss $\phi(x,y,z)$ this yields the dissipated energy $W_\mathrm{diss}$ (see Eq.~\ref{Eq:Wdiss}). In the 1D periodic structure three main contributions may be identified: The elastic energy due to $\pm \sigma_\mathrm{zz}$ pressures on the front and back side of the optical grating (i.e. the upper grating region), the elastic energy due to the $\pm \sigma_\mathrm{xx}$ pressures on the side walls of the optical grating and the elastic energy caused by $\pm \sigma_\mathrm{xx}$ pressures on the side walls of the supporting structure. Cross terms account for about $5\%$ of the total elastic energy. The field distribution in the structure and therewith the stress tensor component depends on the geometric parameters of the grating structure. Thus, before calculating thermal noise in the structure, in the following section we will explain how suitable parameters yielding high reflectivity are determined.

\section{Choice of geometric grating parameters}
\label{sec:params}

\noindent As shown in Fig.~\ref{fig:sketch}, five parameters characterize the structure of a grating reflector: grating period $\Lambda$, width $w_\mathrm{g}$ and depth $h_\mathrm{g}$ of the optical grating as well as width $w_\mathrm{s}$ and depth $h_\mathrm{s}$ of the support structure. We utilize the rigorous coupled wave analysis (RCWA) \cite{Moh1981}, a standard tool to solve Maxwell's equations in periodic structures for the computation of reflectivity and explore 
how the reflectivity depends on the grating parameters. The basic requirements for suitable parameter sets are: a high reflectivity, low field enhancement inside the structure to minimize virtual pressure and possibly compact structures to minimize the elastic deformation energy. Thus, we choose high reflectivity configurations employing low-Q optical resonances with low field enhancement \cite{Kar2012} and minimize the total depth $h_\mathrm{g}+h_\mathrm{s}$. As mentioned above, the supporting structure's task is to optically decouple the optical grating from the substrate. The penetration depth of light into the support increases with decreasing refractive index contrast between the optical grating and the support structure. The index contrast, in turn, is determined by the width $w_\mathrm{s}$ of the supporting structure. Hence $w_\mathrm{s}$ is an important parameter for Brownian thermal noise and is used as a free parameter in the following discussions. \newline
\begin{figure}[tb]
	\centering
		\includegraphics[width=0.5\textwidth]{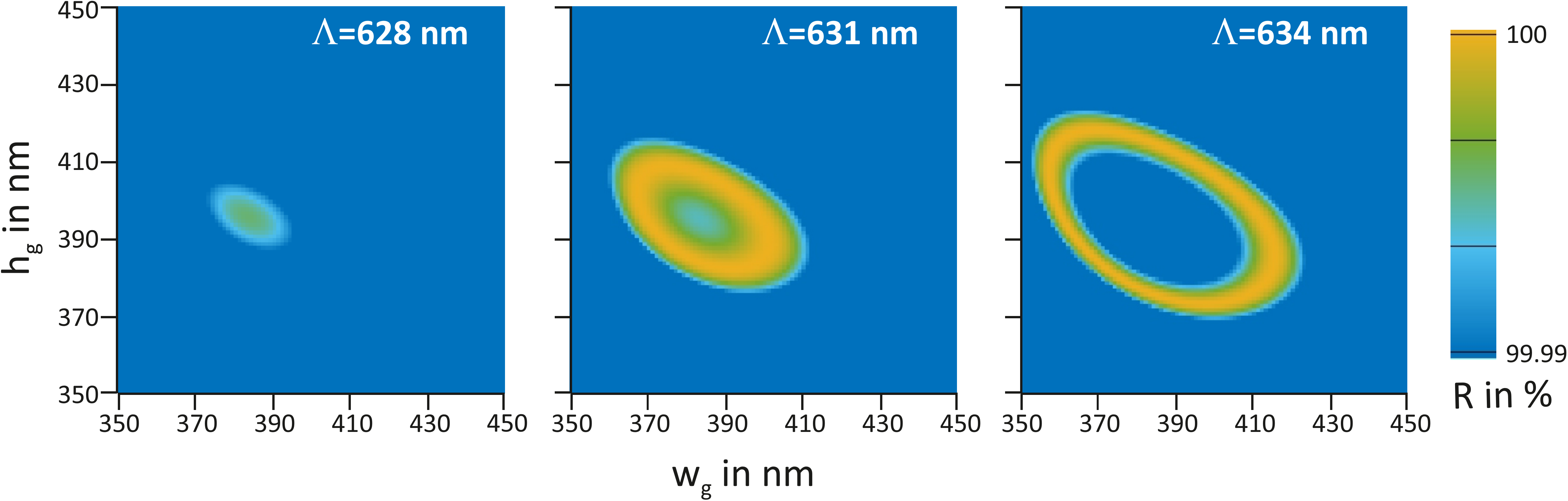}
	\caption{Calculated high reflectivity parameter range $\left\{w_\mathrm{g},h_\mathrm{g}\right\}$ for different periods. A supporting structure width $w_\mathrm{s}$ of 40\,nm was utilized. The incident light has a wavelength of 1550\,nm, an incidence angle of 0$^\mathrm{\circ}$ and transverse-magnetic polarization.}
	\label{fig:reflectance}
\end{figure}
\begin{table}[b]
\begin{center}
	\caption{Structure parameters for $R>99.99\%$. In the $h_\mathrm{g}$ - $w_\mathrm{g}$ plot (see Fig.~\ref{fig:reflectance}) the parameter sets represent the center of the highly reflective parameter space $\left\{w_\mathrm{g},h_\mathrm{g}\right\}$ for a wavelength of 1550\,nm, normal incidence and TM-polarization. Additionally, the structure parameters utilized by Heinert \textit{et al.} \cite{Hei2013} are displayed.}\vspace{0.5cm}
  \begin{tabular*}{\columnwidth}{l@{\extracolsep\fill}ccccc}
		\hline\hline
    $w_\mathrm{s}$ in nm & $\Lambda$ in nm & $h_\mathrm{s}$ in nm & $w_\mathrm{g}$ in nm & $h_\mathrm{g}$ in nm\\ \hline
    40 & 631 & 640 & 385 & 395 \\
		60 & 633 & 630 & 385 & 392 \\
		80 & 637 & 620 & 382 & 390 \\
		100 & 642 & 620 & 384 & 384 \\
		120 & 649 & 630 & 384 & 379 \\
		140 & 660 & 670 & 386 & 371 \\
		160 & 675 & 750 & 385 & 361 \\
		172\footnote{Structure used by Heinert \textit{et al.} \cite{Hei2013}.} & 688 & 800 & 388 &350\\
		180 & 701 & 920 & 391 & 345 \\
		200 & 739 & 1280 & 397 & 335 \\
    220 & 776 & 2200 & 409 & 322 \\  \hline
  \end{tabular*}
	\label{tab:Parameters}
\end{center}
\end{table}
\noindent For a given $w_\mathrm{s}$, the size of the parameter space $\left\{w_\mathrm{g}, h_\mathrm{g}\right\}$ providing high reflectivity depends on the grating period $\Lambda$. Its shape and position is determined by the complex interplay of two Bloch modes propagating in the optical grating. This mechanism for high reflectivity is discussed in detail in the works by Lalanne \textit{et al.} \cite{Lal2006} as well as by Karagodsky \textit{et al.} \cite{Kar2010}. Fig.~\ref{fig:reflectance} illustrates the range $\left\{w_\mathrm{g},h_\mathrm{g}\right\}$ calculated with RCWA for three different periods. In order to achieve large fabrication tolerances, the high-reflectivity range of $w_\mathrm{g}$ and $h_\mathrm{g}$ has to be maximized. As illustrated in Fig.~\ref{fig:reflectance}, the size of the relevant parameter range grows with increasing grating period. However, for large grating periods the high-reflectivity domain in the $w_\mathrm{g}-h_\mathrm{g}$ plain degenerates to a ring which is detrimental in terms of fabrication tolerances if the reflectivity drops below the target value inside the enclosed area. For each target reflectivity $R$, which is typically $R\geq99.99\%$ \cite{Kro2015} there exists an optimal period which maximizes the size of the simply connected high-reflectivity area. With $R\geq99.99\%$ the optimal period for the configuration investigated in Fig.~\ref{fig:reflectance} is 631\,nm. The optimal working point is then located in the center of the area obeying $R\geq99.99\%$. The depth of the supporting structure $h_\mathrm{s}$ does not substantially influence the reflectivity distribution within the parameter range $\left\{w_\mathrm{g}, h_\mathrm{g}\right\}$. To achieve structures as compact as possible, at the end of the optimization process the minimal $h_\mathrm{s}$ for $R\geq99.99\%$ may be chosen. Following this strategy, the optimal parameters in dependence of support structure widths $w_\mathrm{s}$ were determined. The resulting values are shown in Table~\ref{tab:Parameters}. It is noteworthy that enhancing $w_\mathrm{s}$ from 40\,nm to 220\,nm increases $h_\mathrm{s}$ by a factor of 3.4 whereas the other parameters change by less than 20\%.

\section{Results and Discussion}
\label{sec:results}
\begin{table}[tb]
\begin{center}
	\caption{Material parameters for silicon. $\phi_\mathrm{grat}$ is the mechanical loss of the grating structure, $\rho$ the density, $Y$ the Young's modulus and $\sigma$ the Poisson ratio.}\vspace{0.5cm}
  \begin{tabular*}{\columnwidth}{l@{\extracolsep\fill}cccc}
		\hline\hline
     & $T=300$\,K & $T=10$\,K \\ \hline
   $\phi_\mathrm{grat}$ & $5\times10^{-5}$ \cite{Yas2000} & $1\times10^{-5}$ \cite{Mam2001} \\
	$\rho$ in $\mathrm{kg/m^3}$&  \multicolumn{2}{c}{2331}\\
		$Y$ in GPa & \multicolumn{2}{c}{130 \cite{Wor1965}}  \\
    $\sigma$ & \multicolumn{2}{c}{0.28 \cite{Wor1965}}\\  \hline
  \end{tabular*}
	\label{tab:material}
\end{center}
\end{table}
\begin{figure}[b]
	\centering
		\includegraphics[width=0.50\textwidth]{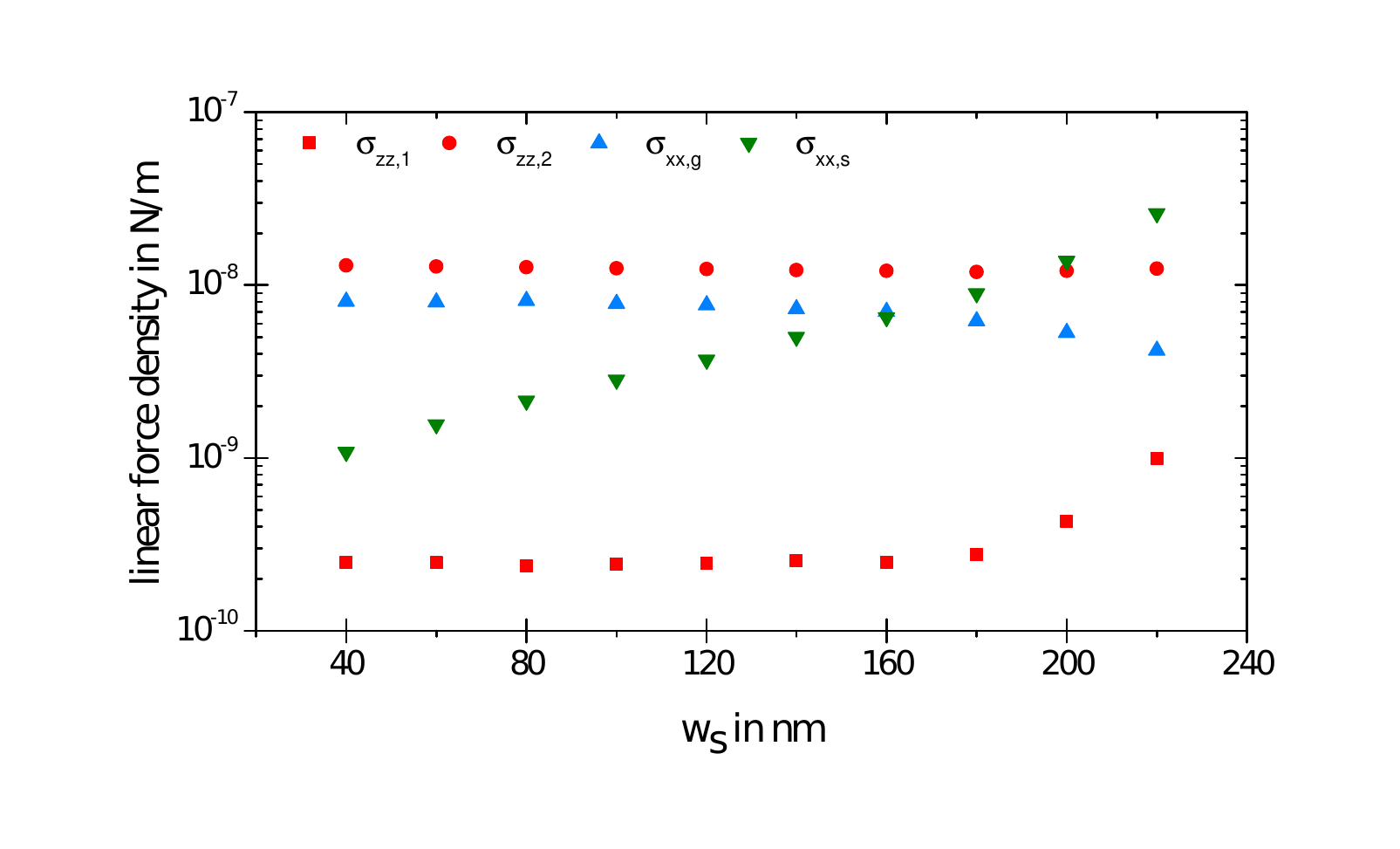}
	\caption{Linear density of virtual forces (per unit length in $y$-direction) at the surface.}
	\label{fig:force}
\end{figure}

\begin{figure}[tb]
	\centering
		\includegraphics[width=0.50\textwidth]{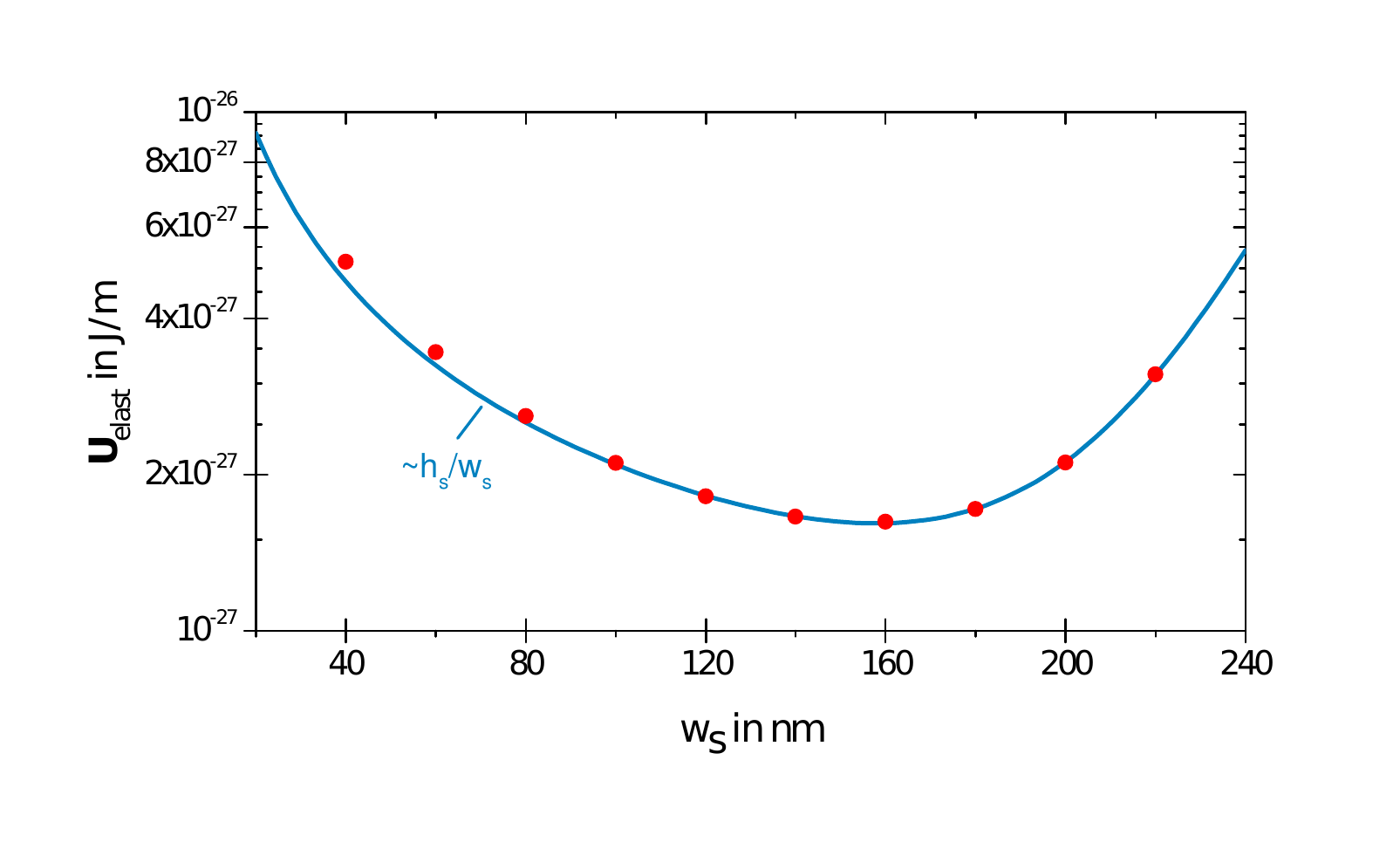}
	\caption{Linear elastic energy density $U_\mathrm{elast}$ (per unit length in $y$-direction) in dependence of the support grating width using a frequency of 100\,Hz.}
	\label{fig:energy}
\end{figure}
\noindent With the grating parameters shown in Table~\ref{tab:Parameters} the stress tensor, the pressure and the resulting force $F_0$ at the grating surface were calculated. The computation of the elastic stress distribution within the grating structure was performed with the finite element tool COMSOL \cite{comsol}. All calculations refer to an incident light power of 1\,W. The power determines the absolute values of the forces and of the elastic energy but it has no influence on the thermal noise amplitude \cite{Dem2015, Tug2017}. The related material parameters are illustrated in Tab.~\ref{tab:material}. Fig.~\ref{fig:force} shows the contributions of the different interfaces to the total force. The colors of the data points correspond to the colors used in Fig~\ref{fig:sketch}. For small $w_\mathrm{s}$ the force at the back side of the optical grating dominates the contributions from the other interfaces. A very similar situation was found by Heinert \textit{et al.} \cite{Hei2013}. There, the magnitude of the force at the front side is by a factor of 54 smaller than the force at the back side. Our calculations reveal a factor of 57. The dominance of the forces at the back side are a consequence of the $E_\mathrm{z}$ distribution in the structure (see Fig.~\ref{fig:Tshapefield}) which is enhanced at the back side of the optical grating.\newline 
In $x$-direction, for small $w_\mathrm{s}$ the optical grating contributes more to the force than the support structure, because the electromagnetic field merely penetrates into the support structure. With increasing $w_\mathrm{s}$ the refractive index contrast between optical grating and support structure decreases. As a result, the field is increasingly pulled into the support structure and the field in the upper region of the support structure increases.
\begin{figure}[b]
	\centering
		\includegraphics[width=0.50\textwidth]{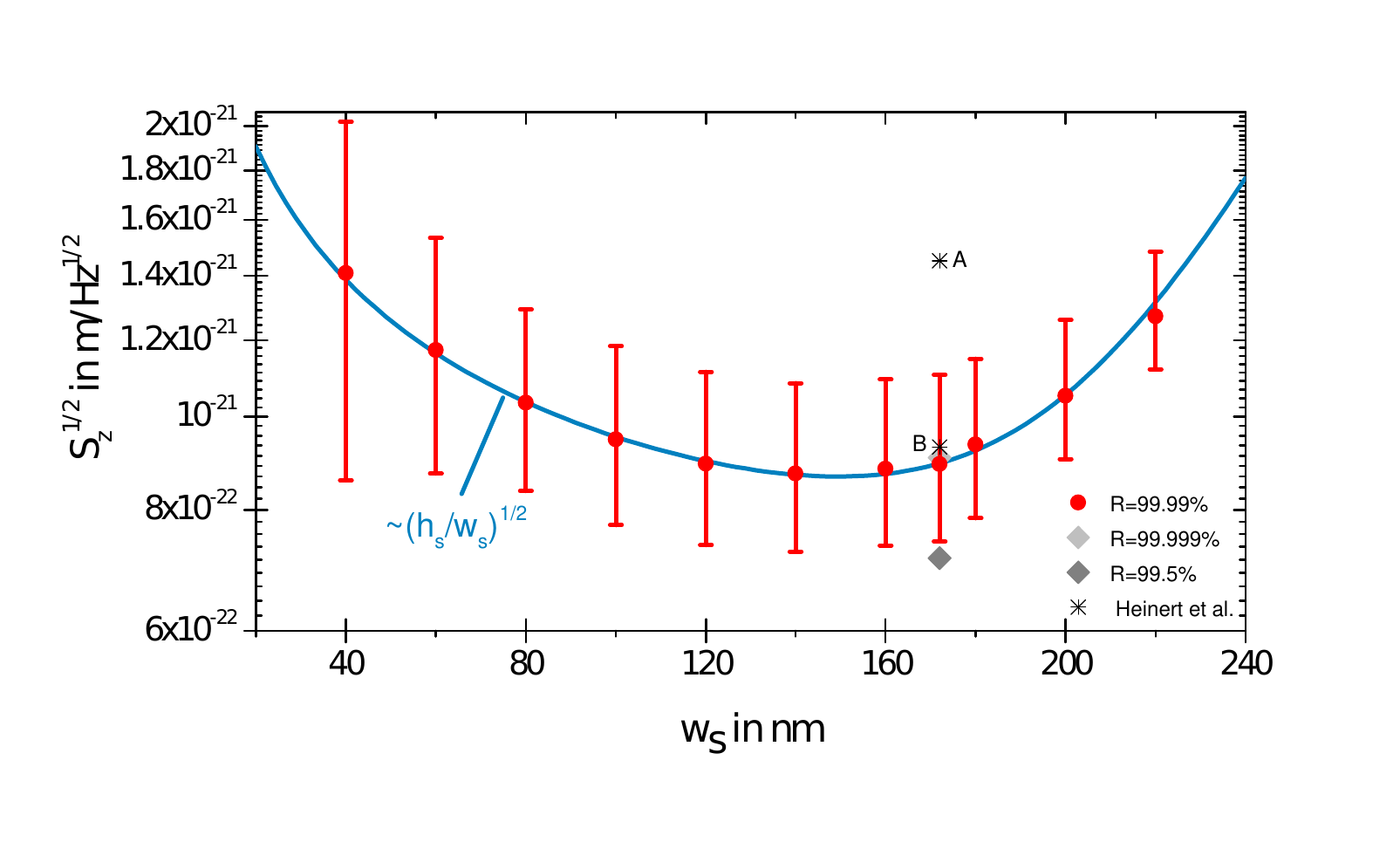}
	\caption{Brownian noise amplitude at a frequency of 100\,Hz and a temperature of 300\,K and a beam radius of 9\,cm resulting from the elastic energies shown in Fig.~\ref{fig:energy}. In addition, datapoint A marks the calculation result by Heinert \textit{et al.} \cite{Hei2013}. For comparison, point B was calculated by applying Maxwell's stress tensor to the same grating parameters. The noise amplitudes for $R>99.5\%$ and $R>99.999\%$ are shown to illustrate the impact of the reflectivity requirement.} 
	\label{fig:noiseamplitude}
\end{figure}
Fig.~\ref{fig:energy} shows the elastic energy $\textit{U}_\mathrm{elast}$ stored in one grating ridge for high reflectivity configurations with different $w_\mathrm{s}$. Here, we refer to the linear elastic energy density per unit length in $y$-direction (compare Fig.~\ref{fig:sketch}): 
\begin{equation}
\textit{U}_\mathrm{elast}=\int\epsilon\ \mathrm{d}x\mathrm{d}z.
\end{equation} 
A frequency of 100\,Hz was used. Fig.~\ref{fig:energy} demonstrates that the ridge behaves like a loaded one-dimensional beam with
\begin{equation}
\textit{U}_\mathrm{elast}=\frac{kx^2}{2}=\frac{F^2}{2k}\propto\frac{h_\mathrm{s}}{w_\mathrm{s}}.
\end{equation}
The ratio $\frac{w_\mathrm{s}}{h_\mathrm{s}}$ represents the spring constant $k$. For small $w_\mathrm{s}$ the elastic energy is dominated by the $1/w_\mathrm{s}$ dependence. Reducing $w_\mathrm{s}$ the spring becomes softer and more elastic energy can be stored. For large $w_\mathrm{s}$ the thickness of the supporting structure needs to be increased to impede light from coupling to the substrate. The increased $h_\mathrm{s}$ again leads to a reduced spring constant and to higher elastic energies. The characteristic dependence on $w_\mathrm{s}$ is also evident in the thermal noise amplitude which is shown in Fig.~\ref{fig:noiseamplitude} for a frequency of 100\,Hz and a temperature of 300\,K. Thermal noise becomes minimal for a support structure width $w_\mathrm{s}$ of about 160\,nm. At cryogenic temperatures Brownian thermal noise is further reduced due to decreased mechanical loss and temperature.\newline
Deviations from the grating design parameters may not only influence the feasible reflectivity but also thermal noise. Therefore, we investigated thermal noise for possible parameter combinations obeying the reflectivity requirement of 99.99\%. To this end, we utilized the parameters given in Table~\ref{tab:Parameters} as working points and performed an error estimation by checking the dependence of thermal noise on the parameters $h_\mathrm{s}$, $w_\mathrm{s}$, $h_\mathrm{g}$ and $w_\mathrm{g}$. Thermal noise remains in the same order of magnitude for all relevant parameter combinations. Deviations of $w_\mathrm{s}$ change the spring constant of the grating and therefore make the largest contributions to changes in thermal noise. For small values of $w_\mathrm{s}$ the reflectivity requirement gives tolerances of about $\pm$10\,nm which are comparable to the values of $w_\mathrm{s}$. That is why small $w_\mathrm{s}$ exhibit the error bars of maximum size. 
\newline 
Finally, we evaluate how thermal noise behaves for slightly different reflectivity requirements of $R>99.5\%$ and $R>99.999\%$. In both cases a $w_\mathrm{s}$ of 172\,nm was chosen. Brownian thermal noise decreases by 20\% for $R>99.5\%$ and increases by 1.6\% for $R>99.999\%$. This variation of thermal noise is a consequence of reduced or enhanced
$h_\mathrm{s}$ which are necessary to achieve $R>99.5\%$ and $R>99.999\%$, respectively.\newline
In comparison to the results by Heinert \textit{et al.} \cite{Hei2013} the calculation with Maxwell's stress tensor yields a thermal noise amplitude which is by a factor 0.61 smaller than the estimate given in the previous work (see Fig.~\ref{fig:noiseamplitude}). The complex field distribution utilized in the present article was treated as a homogenous averaged distribution in \cite{Hei2013}. This leads to the observed deviations of the thermal noise amplitude. 
\newline 
The the electric field distribution and thus also Brownian thermal noise of grating reflectors with 1D periodicity depend on the polarization of the incident light (see Eq.~\ref{Eq:maxwell}). Therefore, in the grating design the polarization dependence has to be carefully taken into account \cite{Dic2017}. With advanced grating concepts 2D periodic structures this dependence can be overcome \cite{Kro2013}.

\section{Conclusion}

\noindent We presented a method to calculate Brownian thermal noise in micro- and nanostructured surfaces. In our approach, computing the Maxwell stress tensor at the dielectric interface leads directly to the mechanical readout variable that is monitored by optical fields. The method is exact and computationally simpler compared to the approximate method developed by Heinert \textit{et al.} \cite{Hei2013} where the fluctuations of each structural part in all possible directions need to be considered separately to calculate the weighing factors. The application of the method to T-shape monolithic grating reflectors reveals the following behavior: for small support widths, the elastic energy is high as the deformation of the support structure in response to the virtual forces becomes high. However, increasing the width requires a detrimental increase of the support structure depth in order to keep the reflectivity high. Therefore, for Brownian thermal noise an optimal $w_\mathrm{s}$ exists that all T-shape grating designs should aim for. The presented method is applicable to arbitrary functional optical surface structures and incident light properties.

\section*{Acknowledgement}
\noindent The authors thank Frank Fuchs (Gitterwerk GmbH, Jena/ Germany) for providing the RCWA code.
S.K. acknowledges the support by the German Research Council (DFG) within research training group "NanoMet -Metrology for complex Nanosystems" (GrK 1952/1).
Y.L. acknowledges the support by the Australian Research Council Future Fellowship.
S.P.V. acknowledges the support by the Russian Foundation for Basic Research (Grant No. 16-52-10069), and National Science Foundation (Grant No. PHY-130586).
This document has LIGO number P1700090.

% Create the reference section using BibTeX:
%\bibliography{GratingNoisenew}

\end{document}